\documentclass[prl,nofootinbib,showpacs,preprint]{revtex4-1}

\usepackage{subfigure}
\usepackage{amsmath}
\usepackage{amsfonts}
\usepackage{amssymb}
\usepackage[dvips]{graphicx}

\usepackage[colorlinks]{hyperref}
\usepackage[usenames,dvipsnames]{color}

\usepackage[applemac]{inputenc}
\usepackage[T1]{fontenc}
\usepackage{amssymb}
\usepackage{amsfonts}
\usepackage{subfigure}
\usepackage{lipsum}
\usepackage{amsfonts} 
\usepackage{braket}
\usepackage{graphicx} 
\usepackage[usenames]{color} 

\usepackage{bbm}

\def\beq{\begin{equation}}
\def\eeq{\end{equation}}
\def\bsp{\begin{split}}
\def\esp{\end{split}}
\def\bea{\begin{eqnarray}}
\def\eea{\end{eqnarray}}
\def\ba{\begin{array}}
\def\ea{\end{array}}

\def\l.{\left.}
\def\r.{\right.}

\def\part{\partial}

\begin{document}
\preprint{UdeM-GPP-TH-16-248}
\title{Gravitationally induced quantum transitions}
\author{A. Landry}
\email{alexandre.landry.1@umontreal.ca}
\author{M. B. Paranjape\footnote{e-mail(corresponding author): paranj@lps.umontreal.ca}} 
\email{paranj@lps.umontreal.ca}
\affiliation{Groupe de physique des particules, D\'epartement de physique,
Universit\'e de Montr\'eal,
C.P. 6128, succursale centre-ville, Montr\'eal, 
Qu\'ebec, Canada, H3C 3J7 }

\begin{abstract}
In this letter,  we calculate the  probability for resonantly induced transitions  in  quantum states due to time dependent gravitational perturbations. Contrary to common wisdom, the probability of inducing transitions is not infinitesimally small.  We consider a system of ultra cold neutrons (UCN), which are  organized  according to the energy levels of the Schr\"odinger equation in the presence of the earth's gravitational field.   Transitions between energy levels are induced by an oscillating driving force of frequency $\omega$. The driving force is created by oscillating a macroscopic mass in the neighbourhood of the system of neutrons.  The neutrons decay in 880 seconds while the probability of transitions increase as $t^2$.  Hence the optimal strategy is to drive the system for 2 lifetimes.  The transition amplitude then is of the order of $1.06\times 10^{-5}$ hence with a million ultra cold neutrons, one should be able to observe transitions.
\end{abstract}


\maketitle


\section{Introduction} 
In this paper we show that gravitationally induced quantum transitions are in fact  easily conceivable, readily calculable and potentially observable. The transitions are to be induced through resonant activation by an external perturbing, oscillatory gravitational source.  We consider a quantum system of ultra-cold neutrons \cite{golub1991ultra}.  The neutrons are organized according to the quantum energy levels of a particle in the earth's (uniform) gravitational field.  The energy levels of this system are well understood and have already been subject to much experimentation \cite{2003LNP...631..355A,Nesvizhevsky:2005ss,Nesvizhevsky:2002ef,Nesvizhevsky:2003ww}  and the system exhibits the first direct observation of gravitational effects on a quantum system.  The energy levels of this system are spaced non-linearly. Therefore, selecting a particular energy difference picks out exactly two levels.   We propose to excite the system with an external, spherical body of mass $M$ placed as near as possible to the system of neutrons. The body will oscillate at the exact frequency corresponding to a transition between two of the energy levels of the system of neutrons.  At resonance, the transition amplitude experiences a significant increase when compared to off-resonance driving frequencies. 

\section{Ultra-cold neutrons in a gravitational field}
Ultra-cold neutrons correspond to a system of neutrons where the temperature is less than a milli-Kelvin.  At these temperatures, the average energy of the neutrons is of the order of $k_bT\approx 8.6\times 10^{-8}eV$.  The neutrons pass between a polished stone base and an absorber/scatterer ceiling, so that they are further distilled such that the energy in their vertical motion is reduced.  Basically, the neutron bounces along the polished base in the $x$ direction, but if it rises up and hits the ceiling, as the ceiling is rough, the neutron is scattered into the $y$ direction and removed from the experiment.  Only those neutrons with very little $z$ energy and momentum and otherwise only $x$ energy and momentum, remain in the experiment.  The vertical energy is so small that the neutrons may only occupy the first four energy levels.  The energy in the vertical motion is bounded by $E_{4}^{0} \approx 4.1\times 10^{-12}eV$.

The neutron quantum system for the $z$ degree of freedom, corresponds to a particle of mass $m_N$, the neutron mass, in a linear gravitational potential $V(z) = m_Ngz$  with an infinite energy barrier at $z=0$.  Therefore the neutron is restricted to $0 \leq z$.   The corresponding Schrodinger equation is:
\begin{equation}
-\frac{\hbar^{2}}{2 m_N} \frac{d^{2} }{dz^{2}} \psi_{E}(z) + m_N g z \psi_{E}(z)=E \psi_{E}(z).
\label{1}
\end{equation}
The normalizable solutions are simply Airy functions, $Ai(\frac{z}{z_{0}}-\alpha )$ is a eigenfunction with energy $m_Ngz_0\alpha$ where $z_0^3=\frac{\hbar^2}{2gm_N^2}$.    Because of the infinite energy barrier at $z=0$, our wave functions must vanish there.  The Airy functions are non-zero for positive argument and have an infinite set of discrete zeros for negative argument at $z/z_0=-\alpha_n$, $n=1,2,3\cdots$.  Therefore the energy eigenfunctions which satisfy the boundary condition are simply $\psi_n(z)={\cal N}_nAi(\frac{z}{z_{0}} -\alpha_{n} )$ where ${\cal N}_n$ is the appropriate normalization \cite{vallee2004airy}:
\begin{equation}
{\cal N}_n=\frac{1}{\sqrt{z_{0} \int_{-\alpha_{n}}^{\infty} Ai(\eta)^{2}d\eta}}= \frac{1}{\sqrt{z_0} Ai'\left(-\alpha_{n} \right)}
\label{2}
\end{equation}
The energy of this eigenstate is of course $m_Ngz_0\alpha_n$.
The $\alpha_n$ are known numerically to arbitrarily high accuracy, however, the Bohr-Sommerfeld approximation \cite{Nesvizhevsky:2003ww,Pedram:2012np} is surprisingly accurate and useful as  it yields an analytic formula:
\begin{eqnarray}
E_{n} =\left(\frac{9 m_N h^{2}g^{2}}{32} \left(n - \frac{1}{4}\right)^{2}\right)^{1/3}\times 10^{-12}eV=1,69 \left(n-\frac{1}{4}\right)^{2/3}peV
 \label{4}
\end{eqnarray}

\section{Gravitational perturbation}
The gravitational perturbation that we imagine the system is subjected to, corresponds to the effect of a macroscopic mass, $M$,  brought as close as possible to the system of neutrons, and subjected to oscillatory motion at exactly the frequency corresponding to a resonance.  It is simplest to imagine the mass $M$ as a spherical body, which is brought to a position $\zeta$ on the $z$ axis, above the system of neutrons.  Its height varies as a $\zeta(t)=\zeta_{0}+\Delta\zeta cos(\omega t)$.  The distance from a neutron at position $z$ is of course $\zeta(t)-z$.  Then the perturbing potential is:
\bea
 W(t,z) =\frac{G m_N M}{\zeta(t)-z}\approx W_{1}(z)-\Delta\zeta W_{2}(z) cos(\omega t) 
\label{7}
\eea 
Where $W_{1}(z)=\frac{G m_N M}{(\zeta_{0}-z)}$ and $W_{2}(z)=\frac{G m_N M}{(\zeta_{0}-z)^{2}}$.  The first term can be treated by  time-independent perturbation theory while the second term needs the time-dependent theory.  The relevant time dependent perturbation theory is that which computes the probability of transitions between two isolated, discrete levels.  As we have noted, the energy differences, and hence the relevant frequencies between any two levels are distinct, as the energy spectrum is non-linear.  Therefore the relevant transitions will be essentially restricted to a two level system.  With an oscillatory driving force, the system in \cite{PhysRevD.81.065019} experiences Rabi oscillations \cite{PhysRev.55.526} and the neutron moves sinusoidally  from one state to the other.  However, in our case, the perturbation is not the same as what is required for Rabi oscillations.  Indeed, we will find that the proper approximation to do is to calculate the short time transition rate between the two levels.  Such a calculation neglects the probability of the transition back to the original state, which is valid if very few transitions occur.  Then, the fact that transitions have occured will be observable only if there are many neutrons available.  We will see that this is indeed the case. 

The first term of (\ref{7}) is to be treated by time-independent perturbation theory.  This will give perturbed static energy levels and eigenfunctions.  The second term will in principle provoke transitions between these perturbed levels.  Since the time dependent perturbation is already very small, it will not be necessary or even consistent to take into account any corrections to the energy levels due to the static perturbation.  Thus we will simply disregard the static perturbation.

For the time-dependent term of equation \eqref{7}, assuming the driving force is started at $t=0$,  the probability of transition between two levels,  $n$ and $m$ is given by
 \bea
P_{nm}(\omega, t)= \frac{ \Delta\zeta^{2}}{\hbar^{2}} \left| \int_0^tdt'\, \left\langle \psi_{m}(z,t')\left| W_{2}(z) cos(\omega t')\right|  \psi_{n}(z,t')\right\rangle \right| ^{2}.
\label{11} 
 \eea
Using the expression for $W_2(z)$, and with the notation $\omega_{mn}=(E_m-E_n)/\hbar$, we get
\beq
P_{nm}(\omega, t)=\left| \frac{G m_N M\Delta\zeta}{\hbar} \left[\int_{0}^{t} dt^{'} exp\left(i \omega_{mn} t^{'} \right) cos(\omega t^{'}) \right]\left[ \int_{0}^{\infty} dz \frac{\psi_{m}(z) \psi_{n}(z)}{\left(\zeta_{0}-z\right)^{2}} \right] \right| ^{2}
\label{12}
 \eeq
 The time integral is trivially done, and it is well  known that on resonance, when $\omega=\omega_{mn}$ it gives a factor
 \beq
\frac{1}{4} \left[\left(\frac{\sin \left( \omega_{mn} t \right) }{\omega_{mn}} \right)^{2} + t \left(\frac{\sin \left(2 \omega_{mn} t \right) }{\omega_{mn}} \right) + t^{2} \right].
 \eeq
This expression is only valid for  short times, as long as no appreciable amount of transitions have been made.  If the level to which transitions are made begins to be macroscopically occupied, then we must take into account the transition induced back to the original level.  The expression Eqn. \eqref{12} does not take into account return transitions, which are being neglected.  The time dependence is essentially parabolic with a slight oscillation about the parabola.  The over all transition rate is controlled by the spatial matrix elements in Eqn. \eqref{12}.  The perturbation that we have considered does not give rise to Rabi type oscillations as considered in \cite{PhysRevD.81.065019}, our perturbation is somewhat different.   

The time-independent integrals correspond to:
\bea
I_{2} \left( \alpha_{m}, \alpha_{n}\right)&=& \int_{0}^{\infty} dz \frac{1}{\left(\zeta_{0}-z\right)^{2}} \psi_{m}(z) \psi_{n}(z)\nonumber\\
&\approx& \frac{1}{\zeta_{0}^{2}} \delta_{mn} + \frac{2}{\zeta_{0}}{\cal N}_m {\cal N}_n\left(\frac{z_{0}}{\zeta_{0}} \right)^{2} \int_{0}^{\infty} dy y  Ai(y -\alpha_{m} ) Ai(y -\alpha_{n} )
\label{14}
\eea
The integral can be found in \cite{vallee2004airy}, we get
\beq
\int_{0}^{\infty} dy y  Ai(y -\alpha_{m} ) Ai(y -\alpha_{n})=\frac{-2}{(\alpha_m-\alpha_n)^2}Ai'(\alpha_m)Ai'(\alpha_n)
\eeq
and substituting for the normalization from Eqn. \eqref{2}, the derivatives of the Airy function nicely cancel and we find
\beq
I_{2} \left( \alpha_{m}, \alpha_{n}\right)\approx \frac{1}{\zeta_{0}^{2}} \delta_{nm} -\frac{4z_0}{\zeta^3_{0}} \frac{1}{(\alpha_m-\alpha_n)^2}
\eeq
Therefore we find, considering the case $m\ne n$, 
\beq
P_{nm}(\omega, t)=\left( \frac{G m_N M\Delta\zeta}{2\zeta_{0}^{2}\hbar}\right)^2\left(\frac{4z_0}{\zeta_{0}} \frac{1}{(\alpha_m-\alpha_n)^2}\right)^2 \left[\left(\frac{\sin \left( \omega_{mn} t \right) }{\omega_{mn}} \right)^{2} + t \left(\frac{\sin \left(2 \omega_{mn} t \right) }{\omega_{mn}} \right) + t^{2} \right]\label{20}
\eeq

\section{Numerical values and experimental possibilities}
The numerical value of the pre-factor in Eqn. \eqref{12} depends on experimental choices.  We will take the most favourable  values imaginable $M = 10$ kg, a sphere of gold will have just under 5 cm radius, hence we can take $\zeta_{0}=5$ cm and $\Delta\zeta = .5$ cm just to find the order of magnitude of the pre-factor.  Then with the values for $G=6.67\times10^{-11}$ Nt m$^2$/kg$^2$, $\hbar=1.054\times 10^{-34}$ Joule sec, $m_N=1,67 \times 10^{-27}$ kg,  $z_{0}=5,874 \times 10^{-6}$ m (which is a characteristic length for the system of neutrons),  and for transitions between the 1st and 2nd energy levels, for which $\Delta E=.493\,\, peV$ and $(\alpha_2-\alpha_1)=1.64$  we get:
\beq
\left( \frac{G m_N M\Delta\zeta}{2\zeta_{0}^{2}\hbar}\right)^2\left(\frac{4z_0}{\zeta_{0}} \frac{1}{(\alpha_m-\alpha_n)^2}\right)^2=3.43\times 10^{-12}{\rm sec}^{-2}
\eeq
This seems to be very small, however, in principle, we can drive the system for a long time.  The neutron lifetime $\tau$ is about $\tau=880$ seconds, that is the number of neutrons diminishes exponentially
\beq
N=N_0e^{-t/\tau}.
\eeq
But the probability of transition increases quadratically with time with an oscillatory variation that we can neglect.  Hence the function to maximize is $t^2e^{-t/\tau}$ which occurs at $t=2\tau$.  Thus with an initial, large number of neutrons, $N_0$, we can optimally pump the system for twice  the lifetime.    This yields a factor of $(2\times 880)^2=3.10 \times 10^6$sec$^2$ which gives a transition probability of $1.06\times 10^{-5}$.  Thus if the initial number of neutrons is say  $N_0=p\times e^2\times 10^5$, where $p$ an integer could be up to 100, then we will have $p\times e^2\times 10^5/e^2=p\times 10^5$ neutrons left, after two lifetimes.  Then this implies that we would have, on average, induced $p$ transitions of neutrons.   

The experimental set up for observing the transitions should be quite straightforward.  We simply imagine the neutrons captured in an open box with sides of height $\langle \psi_1|z|\psi_1\rangle=\frac{ 2\alpha_1 z_0}{3}\approx 9.16$ microns.  Then  neutrons in the first  energy level will be confined inside the box but any neutrons that have been promoted to higher levels, will of course not be trapped inside the box, and will  fall over the edge and outside.  Observing any neutrons outside would be proof that the gravitational perturbation has provoked transitions.

\section{Conclusion}
Ultra cold neutron sources exist in many places, two of the most important ones are at the Institute Langevin-Laue, Grenoble, France \cite{Piegsa:2014kwa}and at the Los Alamos National Laboratory, Los Alamos, US \cite{Saunders}.  Ultra cold neutrons densities of the order of 50/cm$^3$ are easily obtained in volumes that are of the order of a cubic metre.  Therefore obtaining a system with a million cold neutrons is not absurdly unimaginable.  However UCN are defined as those that can be contained in a material bottle, and this only requires their kinetic energies to be in the range of $10^{-9}eV$, which is very much more than what we can tolerate.  We require extremely cold neutrons, with vertical kinetic energies of the order of $10^{-12}eV$. The $Q$ Bounce experiments \cite{Jenke2009318}, now done almost 10 years ago, consisted of about 4500 sufficiently ultra cold neutrons.  It does not seem impossible to get the number of sufficiently cold neutrons to observe the induced transitions.  Observation of gravitationally induced transitions in a quantum system is an important, fundamental physical property of the interplay of gravitation and quantum mechanics.

\section{Acknowledgments } 
We thank  NSERC of Canada, Biothermica Corporation and the Sibylla Hesse Foundation for their financial support and Georges Azuelos and Richard MacKenzie for useful discussions.


\bibliographystyle{apsrev}
\bibliography{ref}

\end{document}